\documentclass[twocolumn,aps,prb]{revtex4}

\usepackage{ae}
\usepackage[T1]{fontenc}
\usepackage[ansinew]{inputenc}
\usepackage{amsmath}
\usepackage{amssymb}
\usepackage{graphicx}
\usepackage{color}
\usepackage[colorlinks]{hyperref}
\usepackage{subfigure}

\begin{document}

\title{Selective self-excitation of higher vibrational modes of graphene nano-ribbons and carbon nanotubes through magnetomotive instability}
\author{A. Nordenfelt\footnote{Corresponding author. E-mail: anders.nordenfelt@physics.gu.se}}
\affiliation{Department of Physics, University of Gothenburg, SE-412
96 G{\" o}teborg, Sweden}

\begin{abstract}
We demonstrate theoretically the feasibility of selective self-excitation of higher-mode flexural vibrations of graphene nano-ribbons and carbon nanotubes by the means of magnetomotive instability. Apart from the mechanical resonator, the device consists only of a constant voltage source, an inductor, a capacitor, a gate electrode and a constant magnetic field. Numerical simluations were performed on both graphene and carbon nanotubes displaying an overall similar behaviour, but with some differences arising mainly due to the non-linear mechanical bending forces. The advantages and disadvanatges of both materials are discussed.   
\end{abstract}

\maketitle

\section{Introduction}

The interplay between the mechanical and the electronic degrees of freedom of nano sized devices continue to be an active research area on both the experimental and theoretical levels, largely due to the discovery of new high performing materials such as graphene\cite{geim} and carbon nanotubes\cite{iijima}. Interest lies not only in the possible technological applications of such materials but also on a more fundamental level, exemplified for example in the numerous schemes for mechanical ground state cooling that have been proposed in recent years. \\

In two previous papers \onlinecite{anders1, anders2} it was explored theoretically the electro-mechanics of a carbon nanotube that was incorporated as a displacement sensitive resistor in a simple electronic circuit. The two device geometries considered were an RC-circuit under constant current bias and an RLC-circuit under constant voltage bias. It was shown that under the influence of a constant magnetic field the electronic circuit would either pump energy into the mechanical system, possibly leading to an instability and selfexcitation of vibrations, or the opposite, to drain the mechanical oscillation of energy leading to an effective cooling. Remarkably, in most cases, by simply reversing the direction of the magnetic field one could switch between self-oscillation and cooling. On the most basic level, the difference in performance between the two different device geometries consisted in the fact that in the voltage-bias regime the driving voltage could be lowered, but at the expense of having to include a sufficiently large inductance in the system. On a more abstract level, an interesting qualitative difference showed up due to the different dimensions of the corresponding dynamical systems. In an RC-circuit there is essentially only one electronic time scale involved, namely the RC-frequency $\omega_R = 1/(RC)$. In an RLC-circuit, however, we also have the LC-frequency $\omega_L = \sqrt{1/(LC)}$. The consequence of this extra time constant turned out to be a more complicated dependence of the overall performance on the relative magnitudes of the three frequencies $\omega_0$, $\omega_R$ and $\omega_L$, where $\omega_0$ is the mechanical resonance frequency. Most importantly, not only the direction of the magnetic field determined whether there would be pumping or cooling in the system but instead the direction of the magnetic field in combination with the relative magnitudes of $\omega_L$ and $\omega_0$. To be more precise: Suppose that you first fix the direction of the magnetic field so that it causes either pumping or cooling. If you then adjust the LC-frequency so that $\omega_L - \omega_0$ changes sign then the effect is invariably reversed, that is, pumping switches to cooling and vice versa. In this paper we will show that this feauture, more than beeing a mere curiosity, could actually become useful. \\

In the previous work only the fundamental bending mode was taken into account. However, by considering also the higher harmonics a graphene sheet or a carbon nanotube could be viewed as an infinite set of mechanical oscillators with different frequencies. It is also likely that the amplitude of each of these modes would have an effect on the resistance. If we again consider the voltage-biased device, depicted in Figure ($\ref{electronicpic}$), and adjust the LC-frequency so that $\omega_0 < \omega_L <\omega_2$, where $\omega_2$ is the frequency of the second harmonic, by adjusting the direction of the magnetic field so that it pushes the nanotube or graphene towards the gate electrode we could achieve a situation where the second harmonic is self-excited while at the same time the fundamental mode is kept silent (or even cooled down). This could of course be generalized, if you instead want to selectively self-excite the 2n:th mode you adjust the LC-frequency so that $\omega_{2n-2} < \omega_L <\omega_{2n}$ (The reason why only every second mode can come into consideration will become clear later). For previous results on selective excitation of higher vibrational modes see for example Refs \onlinecite{magnus2, fabio}.  The question is now how far you can actually push this scheme in practice. In principle there is no upper limit to the higher mode frequencies, extending for example into the long sought after THz regime. There are however a number of limiting factors. First of all the effect of the Lorentz force declines as we go up into higher modes. Secondly, it is plausible that the sentivity in resistance to higher mode deflections decline at more or less the same rate. The latter fact also has the consequence that the amplitude of the current oscillations generated by the mechanical oscillation is reduced accordingly. Another complication arises from the non-linear forces which become increasingly important at large amplitudes and for higher modes. \\     

A conclusive answer to how high frequency you can reach is of course hard to establish, especially since the quality of the materials is continually enhanced. The aim of this paper is instead to demonstrate through numerical simulations the validity of the basic conjecture, namely the possibility to selectively self-excite higher modes. For each mode considered we will try to assess appropriate boundaries for certain parameters, especially the magnetic field strength. Because of the generality of the scheme there is no reason to exclusively treat either carbon nanotubes or graphene nano-ribbons, instead we will discuss and compare the advantages and disadvantages of both materials.

\section{The electronic circuit}
\begin{figure}[h]
\centering
\subfigure[Sketch of the proposed electronic circuit.]{
\includegraphics[width = \columnwidth]{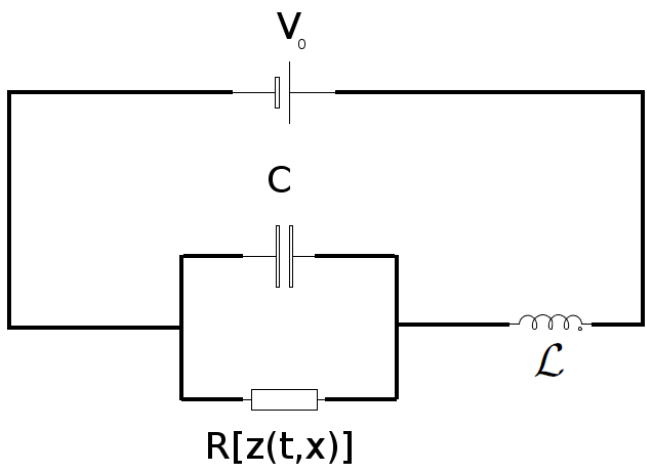}
\label{electronicpic}
}
\subfigure[Sketch of the mechanical resonator comprising the resistor.]{
\includegraphics[width = \columnwidth]{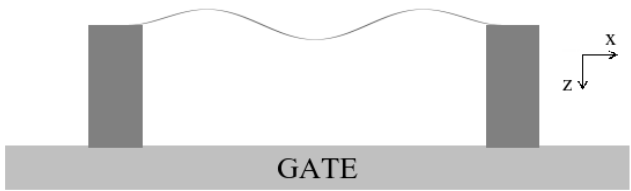}
\label{mechanicpic}
}
\caption{Sketch of the proposed electronic circuit (upper figure). The resistor is comprised of either a graphene sheet or a carbon nanotube suspended over a gate electrode (lower figure).}\label{setup}
\end{figure}
The electronic scheme is depicted in Figure ($\ref{electronicpic}$). The resistor is comprised of either a carbon nanotube or a graphene nano-ribbon that is suspended over a gate electrode. The equations governing the electronic subsystem reads:
\begin{align}\label{electronic}
&C\dot{V} = I - V/R[z(t,x)] \\
&\mathcal{L}\dot{I} = V_0 - V 
\end{align}
The resistance $R$ is now treated as a functional of the total vertical bending shape $z(t,x)$ of the nanotube/graphene where $x$ is the horizontal coordinate, see Figure ($\ref{mechanicpic}$). The selective self-excitation scheme requires that each mechanical bending mode of the material affects the resistance. The way in which this occurs could be outlined as follows. Let the charge per unit length on the resistor be given by the expression
\begin{equation}
n = V_GC_G(z(t,x)),
\end{equation}
where $V_G$ is the gate voltage and $C_G(z(t,x))$ is the gate capacitance per unit length. (This could be thought of as the defining equation for the gate capacitance). For a graphene sheet the following expression for the classical capacitance per unit length was derived in Ref \onlinecite{graphenecap}:
\begin{equation}
C_G = \pi \epsilon_0 \epsilon_r w \left[ 2d\arctan \frac{w}{4d} + \frac{w}{4}\ln \left\{1 + \left(\frac{4d}{w}\right)^2 \right\} \right]^{-1},
\end{equation}
where $w$ is the width of the sheet and $d$ is the distance to the gate electrode. In the limit $w \gg d$ the above expression simplifies to that of a plate capacitor: $C_G = \pi \epsilon_0 \epsilon_r w/d$, and in the limit $w \ll d$ it becomes $C_G = 2\pi \epsilon/\ln(4d/w)$. The latter expression is similar to that for a single walled carbon nanotube: 
\begin{equation}
C_G = \frac{2\pi \epsilon}{\ln(2d/r)},
\end{equation}
which holds in the limit $r \ll d$, where $r$ is the radius of the tube. 
 When we come to discuss the dependence of the resistance per unit length $\zeta$ on the charge carrier concentration $n$ there is a difference. For a single walled semiconducting carbon nanotube, for sufficiently low carrier concentration then to a first approximation we have that\cite{zhou, CNTreview, yury} 
\begin{equation}
\{\zeta(n)\}_{\textrm{SWNT}} \sim \frac{1}{n^2},
\end{equation}
whereas for graphene we expect the following relation to hold\cite{castroneto}
\begin{equation}
\{\zeta(n)\}_{\textrm{Graphene}} \sim \frac{1}{n}.
\end{equation}
In our simulations the total resistance is calculated according to the formula
\begin{equation}
R[z(t,x)] = \int_0^L \zeta(n(C_G(z(t,x)))) \mathrm{dx},
\end{equation}
hence, each segment is treated as an independent resistor whose capacitance only depends on the distance to the gate. This is of course only an approximation. However, we have no reason to believe that the exact quantitative description of the resistance affects the main qualitative conclusions of our analysis, and it will therefore not be discussed at length here. For further details we refer to the references provided.

\section{Carbon nanotubes}

We will first carry out the analysis on a carbon nanotube, which was the mechanical oscillator first considered. The generalization to a graphene nano-ribbon is then straightforward.\\

If we model the carbon nanotube as a doubly clamped elastic beam of length $L$ and include a geometrical nonlinearity term and the external Lorentz force $F_L = HJ$, where $H$ is the magnetic field and $J$ the current through the nanotube, the equation of motion reads
\begin{align}\label{eqmotion}
ES\frac{\partial^4z}{\partial x^4} + \rho A \frac{\partial^2z}{\partial t^2} + \gamma \frac{\partial z}{\partial t} = &\left( \frac{EA}{2L} \int_0^L \left( \frac{\partial z}{\partial x} \right)^2 \mathrm{dx} \right) \frac{\partial^2 z}{\partial x^2} \nonumber \\
& + HJ. 
\end{align}
Here $E$ is the Young's modulus, $\rho$ the mass density, $A$ the cross sectional area, S the area moment of inertia, $\gamma$ the damping coefficient and $L$ the length of the nanotube. If we let $r$ denote the radius of the nanotube we have that $A = \pi r^2$ and $S = \pi r^4/4$. For previous work on the nonlinear dynamics of carbon nanotubes see for example Ref \onlinecite{CNNonDyn}. We chose to treat the problem numerically using a Galerkin reduced-order model, which we truncate at the 6th overtone. Hence
\begin{equation}
z(t,x) = \lambda\sum_{n=0}^{6} u_n(t)\phi_n(x/L)
\end{equation}
Using the notation $\hat{x} = x/L$, the mode shapes are given by the expression
\begin{align}
\phi_n(\hat{x}) = &C_n\{(\sin(k_n) - \sinh(k_n))(\cos(k_n\hat{x}) - \cosh(k_n\hat{x})) \nonumber \\
            &- (\cos(k_n) - \cosh(k_n))(\sin(k_n\hat{x}) - \sinh(k_n\hat{x})\},\nonumber
\end{align}
where $C_n$ are normalization constants chosen so that $\int_0^1\phi_n(\hat{x})^2 d\hat{x} = 1$ and the constants $k_n$ satisfy the equation $\cos(k_n)\cosh(k_n) = 1$. The corresponding vibrational frequencies are given by
\begin{equation}
\omega_n = k_n^2 \sqrt{\frac{ES}{\rho A}}.
\end{equation}
Setting the timescale to $\tau = \omega_0 t$ and projecting equation ($\ref{eqmotion}$) onto each different mode we obtain the set of equations
\begin{equation}
\ddot{u}_n(\tau) + \tilde{\gamma}\dot{u}_n(\tau) + (\omega_n/\omega_0)^2u_n(\tau) = u_n(\tau)\chi_n\Gamma + \alpha_n\frac{LHJ}{K\lambda} 
\end{equation}
where
\begin{align}
&\chi_n = \int_0^1(\phi_n(\hat{x})\phi_n''(\hat{x}))d\hat{x} \\
&\Gamma = \frac{2}{k_0^4}\frac{\lambda^2}{r^2}\int_0^1\left( \sum_{i=0}^{6}u_i(\tau)\phi_n'(\hat{x})\right)^2 d\hat{x}\\
&K = k_0^4ES/L^3 \\
&\alpha_n = \int_0^{1} \phi_n(\hat{x}) d\hat{x}
\end{align}
The constant $Q = 1/\tilde{\gamma}$ will be referred to as the quality factor of the nanotube. We have assumed that the deflection is small enough that the Lorentz force can be considered uniform across the nanotube. In particular this implies that the external force vanishes for n = 1,3,5... so that we only need to include the modes n = 0,2,4,6 into our consideration. In Table ($\ref{CNTpar}$) are listed some of the mode dependent parameters.
\begin{table}[h!b!p!]
\caption{Mode dependent vibrational frequencies $\omega_n$ and integration constants $\alpha_n$ for the first four bending modes with even index of a carbon nanotube.}\label{CNTpar}
\begin{center}
\begin{tabular}{|l||l|l|}
\hline
n & $\omega_n/\omega_0$ & $\alpha_n$ \\ \hline
0 & 1 & 0.83 \\
\hline
2 & 5.4 & 0.36 \\
\hline
4 & 13.3 & 0.23 \\
\hline
6 & 24.8 & 0.17 \\
\hline
\end{tabular}\\
\end{center}
\end{table}
We will now review some of the analytical results that will form the basis of reasoning in this paper. The measure of the sensitivity of the resistance with respect to displacement was expressed by the so called characteristic length, which in order to include higher modes can be generalized in the following way:
\begin{equation}
\lambda_n = -\frac{R}{\partial R/\partial u_n}.
\end{equation}
In the case which we consider, when the sensitivity of the resistance is achieved through electronic doping controlled by a gate electrode, it could be argued that the relative magnitude of the characteristic lengths should be approximately
\begin{equation}
\frac{\lambda_n}{\lambda_m} \approx \frac{\alpha_m}{\alpha_n}.
\end{equation}
We now introduce the mode dependent coupling parameter
\begin{equation}
\beta_n = \alpha_n\frac{HJL}{K\lambda_n}
\end{equation}
and the mode dependent succeptibility function\cite{anders2}
\begin{equation}
S_n = \frac{\omega_n\omega_R(\omega_L^2/\omega_n^2 - 1)}{\omega_n^2(\omega_L^2/\omega_n^2 - 1)^2 + \omega_R^2}.
\end{equation}   
Neglecting non-linear terms and other intermode crosstalk it can be shown that if the product $\beta_nS_n$ is negative the mode becomes unstable approximately when
\begin{equation}
|\beta_nS_n| > 1/Q.
\end{equation}
As mentioned earlier, if for example we assume that $\omega_{2n-2} < \omega_L < \omega_{2n}$ then for a fixed magnetic field it is clear that $\beta_{2n-2}S_{2n-2}$ and $\beta_{2n}S_{2n}$ have opposite signs. If $\beta_nS_n$ is negative then it will also be negative for higher modes but since the averall succeptibility decreases with $n$ it should be possible to find a magnetic field strength where the slight heating effect on the higher modes does not turn into an instability. What we have discussed so far is a more or less straightforward generalization of previous considerations. The main practical complication that arises here is due to the possible interaction between the modes mediated by the non-linear force term. On this issue, the radius of the nanotube is one of the important parameter that we need to take into account. Since the nonlinear-force becomes appreciable only when the amplitude reaches a magnitude of the order of the radius one might think that a large radius is beneficial for our purposes. Indeed, a large radius not only increases the range of amplitude within the linear regime but also increases the spring constant $K$ with the side effect that the thermal noise is reduced, which can be seen from the equipartition relation $1/2k_BT = 1/2K\langle u_0^2\rangle$. But, this comes at the cost of a reduced coupling parameter. If we make the estimates\cite{CNFundMech} $E = 1$~TPa, $\rho = 1.35$g/cm$^3$, $L = 1\mu$m, in Table ($\ref{tubetable}$) we have calculated the key parameters for a number of different radii.
\begin{table}[h!b!p!]
\caption{Effective spring constant, thermal noise and vibrational frequency of a carbon nanotube for three different radii.}\label{tubetable}
\begin{center}
\begin{tabular}{|l||l|l|l|}
\hline
$r$ [nm]& $K$ & $\sqrt{\langle u_0^2 \rangle}$ [nm] at T=1K & $\omega_0$ [$10^9/s$]\\ \hline
0.5 & 2.45$\cdot 10^{-5}$ & 0.75 & 0.15\\
\hline
1.0 & 3.93$\cdot 10^{-4}$ & 0.19 & 0.30\\
\hline
5.0 & 2.45$\cdot 10^{-1}$ & 0.0075 & 1.52\\
\hline
\end{tabular}
\end{center}
\end{table}
As we can see, going from a radius of 0.5 nm to a radius of 5 nm means that $K$ increases by four orders of magnitude, and consequently, the critical magnetic field increases accordingly.\\ 

Apart from the excitation of an overtone, one can also anticipate a certain cooling of the lower harmonics. In case we wish to selfexcite the mode with frequency $\omega_{2n}$ the choise of $\omega_L$ and $\omega_R$ that simultaneously maximizes the excitation succeptibility $S_{2n}$ and the cooling succeptibility $S_{2n-2}$ can be derived straightforwardly:
\begin{align}
& \omega_R = (\omega_{2n}^2 - \omega_{2n-2}^2)/(\omega_{2n} + \omega_{2n-2}) \\
& \omega_L = \sqrt{\omega_{2n-2}\omega_R + \omega_{2n-2}^2}
\end{align}
In our simulations we took these as our default values for the electronic frequencies. We chose to consider a nanotube with radius $r = 1$nm and length $L = 1\mu$m situated at a distance 200 nm from the gate and put $\lambda$=1nm. Numerical simulations were performed using a Runge-Kutta algorithm for different values of $Q$, $\omega_R$, $\omega_L$ and the coupling parameter
\begin{equation}
\beta = \frac{HV_0L}{KR_0\lambda},
\end{equation}
where $R_0$ is the resistance of the unbent nanotube. In each time step the normalized resistance was calculated through numerical integration using 100 steps. The algorithm was implemented in both Matlab and C producing consistent results. 
\begin{figure}[h]
\includegraphics[scale=0.4]{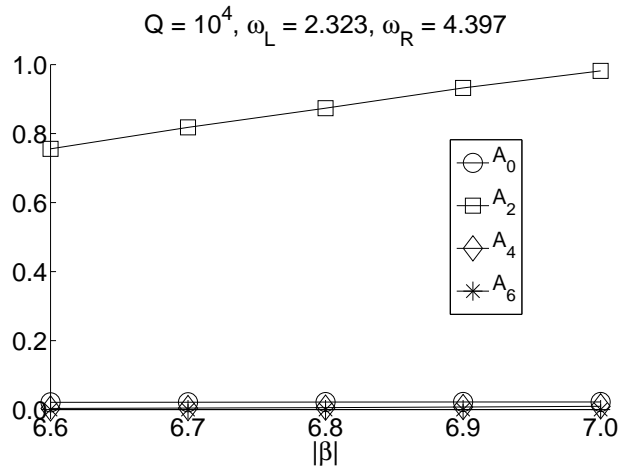}
\caption[]{Saturation amplitudes $A_0$, $A_2$, $A_4$ and $A_6$ for the respective modes $u_0$, $u_2$, $u_4$ and $u_6$, as a result of computer simulations aimed at selective self-excitation of the second harmonic of a carbon nanotube with radius $r = 1$nm and length $L=1\mu$m. The coupling parameter $\beta$ was in the range -6.6 to -7.1 which, assuming a stationary current of 1 $\mu$A, corresponds to a magnetic field strength in the interval 2.6-2.8 T. The other parameters were $Q = 10^4$, $\omega_L= 2.323$, $\omega_R = 4.397$. Initial conditions were 0 in all variables.}\label{sat2CNT}
\end{figure}

\begin{figure}[h]
\centering
\subfigure{
\includegraphics[scale=0.4]{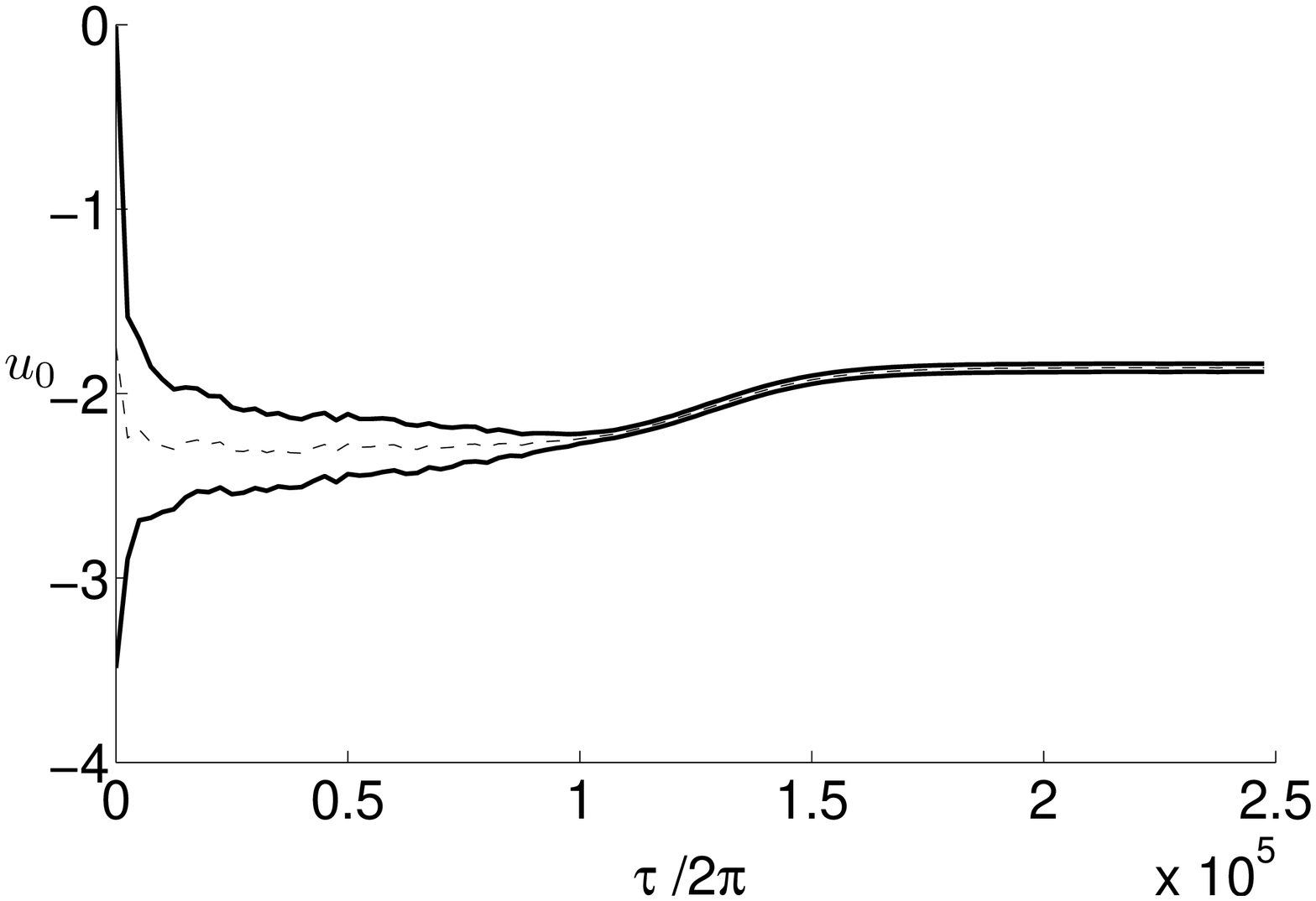}
\label{fig1:subfig1}
}
\subfigure{
\includegraphics[scale=0.4]{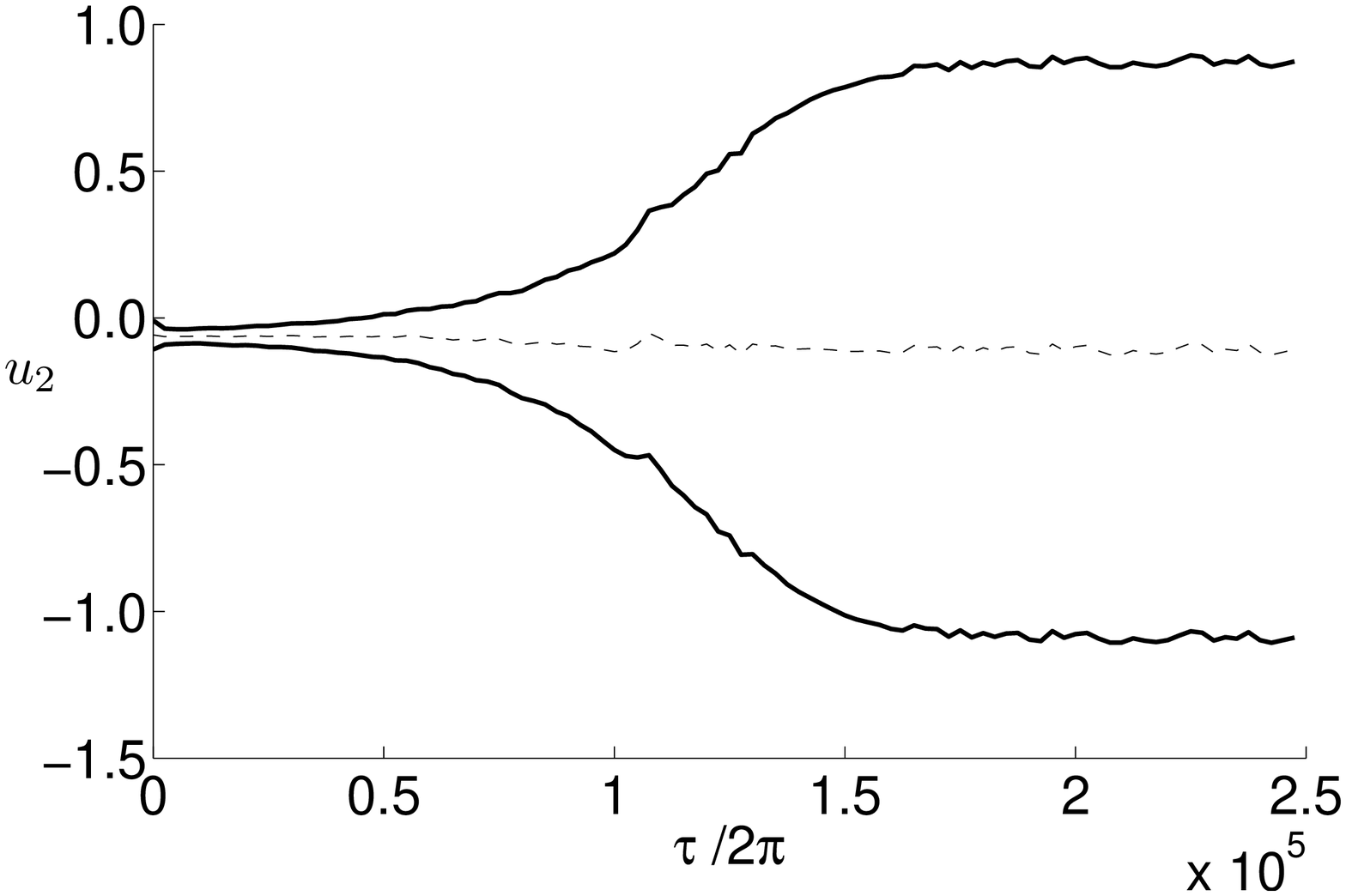}
\label{fig1:subfig2}
}
\caption[]{Time evolution of the envelopes of the rapid oscillations in the fundamental mode $u_0$ (upper figure) and second harmonic $u_2$ (lower figure) as the results of a computer simulation on a carbon nanotube with radius $r = 1$nm and length $L=1\mu$m. The dashed lines mark the displacement averaged over a period of $2\pi$. The parameters were $\beta = -0.007$, $Q = 10^4$, $\omega_L= 2.323$, $\omega_R = 4.397$. Initial conditions were 0 in all variables.}\label{secondnormal}
\end{figure}

In Figure ($\ref{sat2CNT}$) we show the saturation amplitudes $A_0$, $A_2$, $A_4$ and $A_6$ for the respective modes $u_0$, $u_2$, $u_4$ and $u_6$, as a result of computer simulations for a coupling parameter $\beta$ in the range -6.6 to -7.1 which, assuming a stationary current of 1 $\mu$A, corresponds to a magnetic field strength in the interval 2.6-2.8 T. The quality factor was assumed to be $Q = 10^4$ and the electronic frequencies where chosen so as to selectively excite the second overtone. Figure ($\ref{secondnormal}$) shows the envelopes of the rapid oscillations as a function of time of the zeroth and second mode for one of these simulations. As we can see, in the beginning the fundamental mode oscillates with the largest amplitude but later on the second harmonic experiences a rapid boost which causes the magnitude of the timeaveraged displacement in the fundamental mode to decrease. This is because the oscillation increases the effective resistance leading to a reduced Lorentz force. The latter phenomenon was explained and quantified in Refs \onlinecite{anders1, anders2}. The other modes also experience a certain excitation but with an amplitude roughly two orders of magnitude less than that of the second harmonic.  \\
   
\begin{figure}[h]
\includegraphics[scale=0.4]{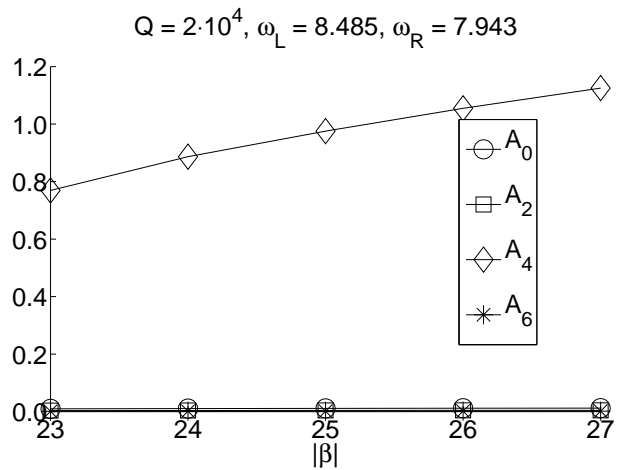}
\caption[]{Saturation amplitudes $A_0$, $A_2$, $A_4$ and $A_6$ for the respective modes $u_0$, $u_2$, $u_4$ and $u_6$, as a result of computer simulations aimed at selective self-excitation of the fourth harmonic of a carbon nanotube of radius $r=1$nm and length $L=1\mu$m. The coupling parameter $\beta$ was in the range -23 to -27 which, assuming a stationary current of 1 $\mu$A, corresponds to a magnetic field strength in the interval 9.0-10.6 T. The other parameters were $Q = 2 \cdot 10^4$, $\omega_L= 8.485$, $\omega_R = 4.397$. Initial conditions were 0 in all variables.}\label{sat4CNT}
\end{figure}

There are a number of ways by which the non-linear forces can cause the scheme to behave in an erratic fashion. If the amplitude of some mode is forced beyond a certain critical value, non-linear forces causes the system either to enter a chaotic state which cannot be tracked numerically, (which was the typical scenario for carbon nanotubes), or to collapse and start all over again, (which was the more common outcome for graphene sheets, to be disussed later). Another effect of the non-linear forces is that the static deflection increases the effective spring constant so that one might end up with a situation where the renormalized frequency of the fundamental mode is suddenly larger than $\omega_L$. Although the latter effect is only an issue when $\omega_L$ is relatively close to $\omega_0$, also when attempting to self-excite higher modes the static deflection induced by the Lorentz force might cause a significant decrease in the effective electro-mechanical coupling so that for quality factors below a certain value there might not be any possibility for self-excitation whatsoever. For example, given a quality factor of $Q = 10^4$ we were not able to find any range of $\beta$ corresponding to a successful self-excitation of the fourth harmonic. However, by increasing the quality factor slightly to $Q = 2 \cdot 10^4$ this was achieved in a number of computer simulations presented in Figure ($\ref{sat4CNT}$). As in the previous case, the saturation amplitude of the other modes was approximately two orders of magnitude less than that of the excited mode. Assuming, as before, a stationary current of 1 $\mu$A the parameter range in this case corresponds to a magnetic field in the interval 9.0-10.6 T.  

\section{Graphene}

We will now carry out the same analysis on a graphene nanoribbon. The elastic and dynamic properties of graphene has been analyzed by several authors\cite{bunchgraph, garciagraph, isacgraph, juanNL}. After a few simplifications, for example that we only need to take into account the vertical streching, the one-dimensional equation of motion of a doubly clamped graphene sheet reads

\begin{align}
&\rho \frac{\partial^2 z(t,x)}{\partial t^2} + \rho \gamma \frac{\partial z(t,x)}{\partial t} - T_0\frac{\partial^2 z(t,x)}{\partial x^2} - \nonumber\\
&-T_1\frac{\partial}{\partial x}\left(\frac{\partial z(t,x)}{\partial x}\right)^3 = P_z(t,x),
\end{align}
where $\rho$ is the area mass density of graphene, $P_z(t,x)$ is the pressure in the vertical direction, $T_0 = (\lambda + 2\mu)\delta$, $T_1 = \lambda/2 + \mu$, $\lambda$ and $\mu$ beeing the Lam\'{e}-parameters and $\delta$ a parameter that quantifies the initial in-plane streching. The important thing to notice is that the nonlinear term is not dependent on the width of the sheet, unlike the carbon nanotube where the radius entered explicitly into the equation. Here, it is instead primarily the length of the nanoribbon that determines the critical amplitude of motion when non-linear forces become dominant. In our simulations we chose the typical parameter values $\lambda$ = 15.55 J/$m^2$, $\mu$ = 103.89 J/$m^2$, $\delta$ = 0.5$\%$\cite{juanNL}, which together with the length $L = 1\mu$m yields a fundamental frequency of the order of 1GHz. For the sake of comparison with the previous simulations on carbon nanotubes we chose the width $w = 10$nm. \\

As before we assume a vertical displacement of the form:  
\begin{equation}
z(\tau,x) = \lambda\sum_{n=0}^{6} u_n(\tau)\phi_n(x)
\end{equation}
where the mode shapes $\phi_n$ are now given by  
\begin{equation}
\phi_n(x) = \sqrt{2}\sin((n+1)\pi x),
\end{equation}
and we set $\lambda$ = 1nm. If we again define $\alpha_n = \int_0^1\phi_n(x) \mathrm{dx}$ we obtain the set of parameters presented in Table ($\ref{graphpar}$).
\begin{table}[h!b!p!]
\caption{Mode dependent vibrational frequencies $\omega_n$ and integration constants $\alpha_n$ for the first four streching modes with even index of a graphene nano-ribbon.}\label{graphpar}
\begin{center}
\begin{tabular}{|l||l|l|}
\hline
n & $\omega_n/\omega_0$ & $\alpha_n$ \\ \hline
0 & 1 & 0.90 \\
\hline
2 & 3 & 0.30 \\
\hline
4 & 5 & 0.18 \\
\hline
6 & 7 & 0.13 \\
\hline
\end{tabular}\label{GRAPHpar}\\
\end{center}
\end{table}

\begin{figure}[h]
\includegraphics[scale=0.4]{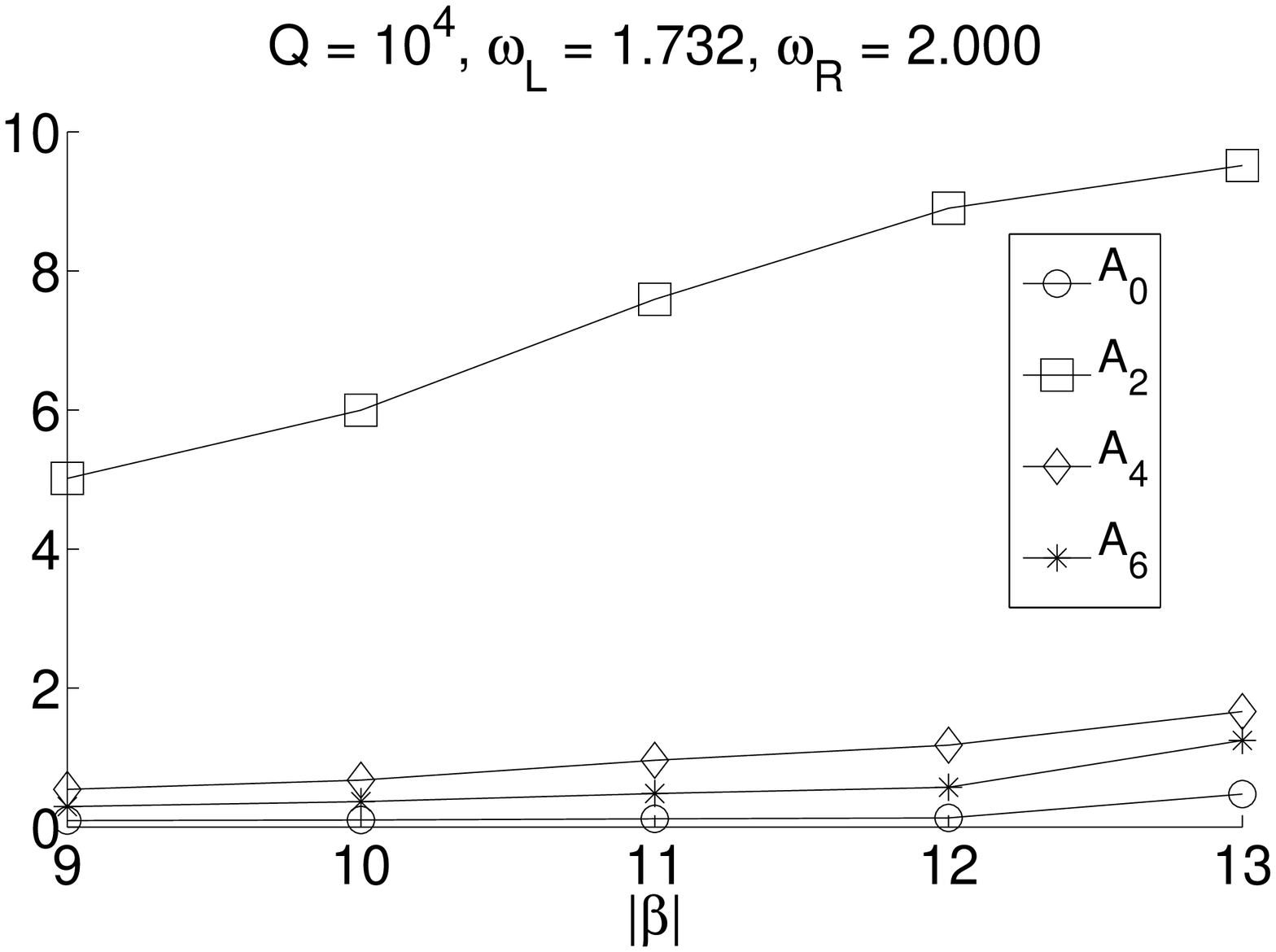}
\caption[]{Saturation amplitudes $A_0$, $A_2$, $A_4$ and $A_6$ for the respective modes $u_0$, $u_2$, $u_4$ and $u_6$, as a result of computer simulations aimed at selective self-excitation of the second harmonic of a graphene nano-ribbon of width $w=1$nm and length $L=1\mu$m. The coupling parameter $\beta$ was in the range -9 to -13 which, assuming a stationary current of 20 $\mu$A, corresponds to a magnetic field strength in the interval 6.6-9.6 T. The other parameters were $Q = 10^4$, $\omega_L= 1.732$, $\omega_R = 2.000$. Initial conditions were 0 in all variables.}\label{sat2graphene}
\end{figure}

\begin{figure}[h]
\includegraphics[scale=0.4]{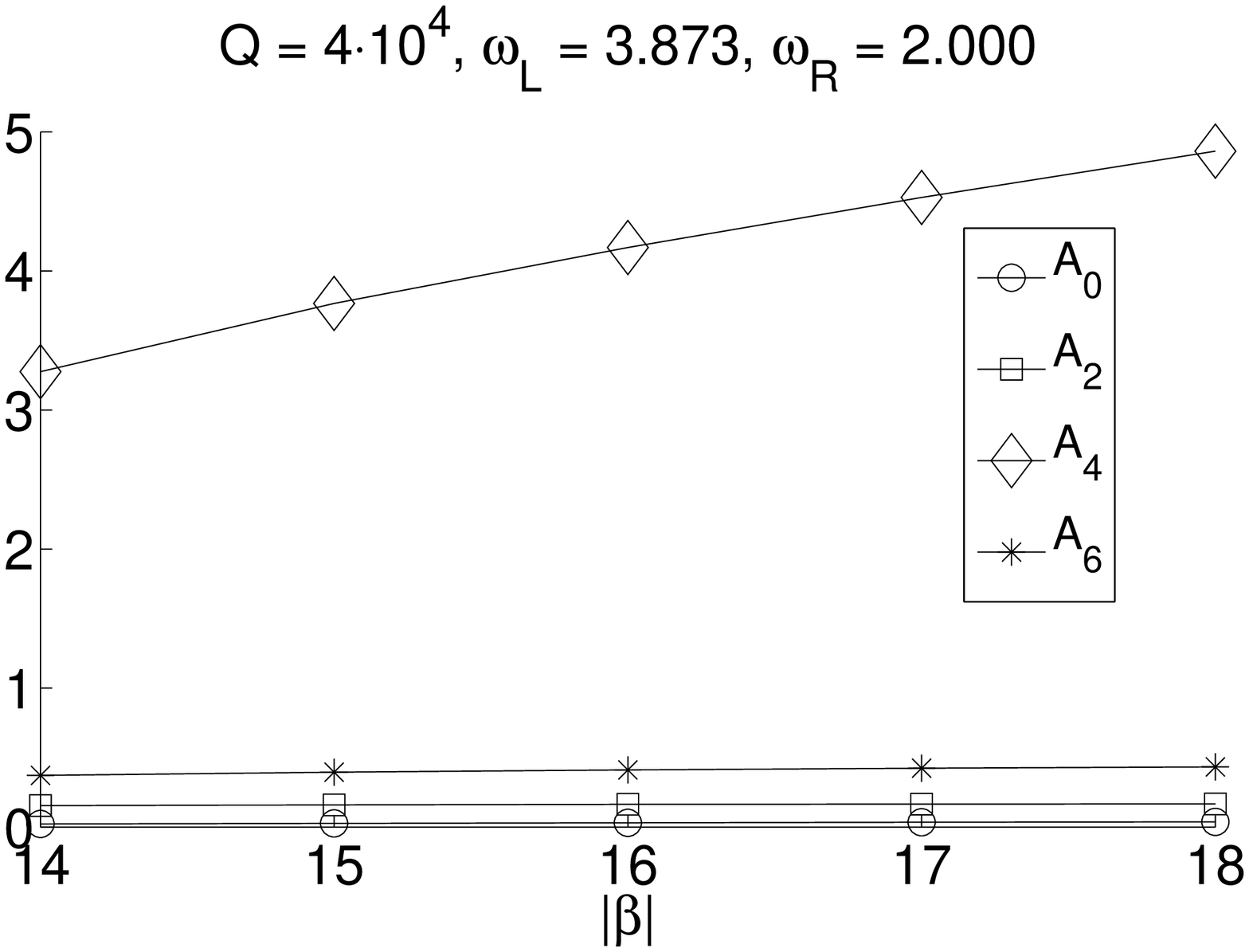}
\caption[]{Saturation amplitudes $A_0$, $A_2$, $A_4$ and $A_6$ for the respective modes $u_0$, $u_2$, $u_4$ and $u_6$, as a result of computer simulations aimed at selective self-excitation of the fourth harmonic of a graphene nano-ribbon of width $w$=1nm and length $L=1\mu$m. The coupling parameter $\beta$ was in the range -14 to -18 which, assuming a stationary current of 20 $\mu$A, corresponds to a magnetic field strength in the interval 10.3-13.2 T. The other parameters were $Q = 4 \cdot 10^4$, $\omega_L= 3.873$, $\omega_R = 2.000$. Initial conditions were 0 in all variables.}\label{sat4graphene}
\end{figure}

As we can see, the $\alpha_n$ decline more rapidly than for carbon nanotubes while the frequencies increase at a slower rate. Both these facts would make a graphene nano-ribbon inferior to a carbon nanotube if it hadn't been for the fact that the amplitude range within the linear regime is greater by almost a factor ten already when comparing a graphene nano-ribbon of length 1$\mu$m to a nanotube of the same length and with radius 1 nm. This can be further extended for graphene by simply making it longer. Numerical simulations were performed, now with the coupling parameter
\begin{equation}
\beta = \frac{HV_0}{AR_0}\frac{t_0^2}{\lambda\rho},
\end{equation}    
where A is the area of the graphene sheet and $t_0 = L\sqrt{\rho/T_0}$. One might think that a small area is beneficial, but on the other hand a larger sheet has a higher current carrying capacity so these two factors are likely to cancel each other out. In Figure ($\ref{sat2graphene}$) we show the saturation amplitudes resulting from a number of computer simulations aimed at exciting the second mode. One obvious difference compared with carbon nanotubes is that there is less of a difference between the saturation amplitude of the excited mode and that of the other modes. For graphene they differ typically by a factor 10 whereas for nanotubes it is almost by a factor 100. The same conclusion can be drawn from the results presented in Figure ($\ref{sat4graphene}$), where the fourth mode is excited.

\section{Concluding remarks}

In this paper we demonstrated theoretically the feasibility of selective self-excitation of the second and fourth harmonic of flexural vibrations of graphene nano-ribbons and carbon nanotubes with quality factors of $10^4$ and upwards. This is accomplished by the means of a constant voltage source, an inductor, a capacitor, a gate electrode and a constant magnetic field. An advantage with graphene is that it allows a higher amplitude of oscillation before non-linear forces spoils the scheme, but, computer simulations indicate that for carbon nanotubes the ratio between the saturation amplitude of the excited mode and that of the other modes is substantially larger, thus allowing a more distinct selective excitation. 

\section{Acknowledgements}
The author wishes to thank Leonid Gorelik and Mats Jonson for useful suggestions and careful proof-reading. The author is though solely responsible for any remaining errors or shortcomings. Financial support from the Swedish research council VR is gratefully acknowledged. 

\bibliography{selective}

\end{document}